\documentclass[journal=jacsat,manuscript=letter]{achemso}
\setkeys{acs}{articletitle = true}
\setkeys{acs}{maxauthors=10}
\setkeys{acs}{etalmode=truncate}

\usepackage{graphicx}
\usepackage{braket}
\usepackage{textcomp}

\title{Impact of electron-phonon scattering on optical properties of CH$_{3}$NH$_{3}$PbI$_{3}$ hybrid perovskite material}

\author{Benoit Galvani}
\affiliation{Aix Marseille Universit{\'e}, CNRS, Universit{\'e} de Toulon, 
IM2NP UMR 7334, 13397, Marseille, France}

\author{Daniel Suchet}
\affiliation{Ecole
Polytechnique, IPVF, Institut Photovoltaïque d\textquotesingle Ile-de-France, 30 RD 128, 91120 Palaiseau, France}
\author{Amaury Delamarre}
\affiliation{Centre de Nanosciences et de Nanotechnologies, Palaiseau, Ile-de-France, France}

\author{Marc Bescond}
\affiliation{LIMMS, CNRS-Institute of Industrial Science, UMI
2820, University of Tokyo, 153-8505 Tokyo, Japan}

\author{Fabienne Michelini}
\affiliation{Aix Marseille Universit{\'e}, CNRS, Universit{\'e} de Toulon, 
IM2NP UMR 7334, 13397, Marseille, France}

\author{Michel Lannoo}
\affiliation{Aix Marseille Universit{\'e}, CNRS, Universit{\'e} de Toulon, 
IM2NP UMR 7334, 13397, Marseille, France}

\author{Jean-Francois Guillemoles}
\affiliation{IPVF, Institut Photovoltaïque d\textquotesingle Ile-de-France, 30 RD 128, 91120 Palaiseau, France}

\author{Nicolas Cavassilas}
\email{nicolas.cavassilas@im2np.fr}
\affiliation{Aix Marseille Universit{\'e}, CNRS, Universit{\'e} de Toulon, 
IM2NP UMR 7334, 13397, Marseille, France}

\begin{document}
\begin{abstract}
We investigate numerically the impact of the electron-phonon scattering on the optical properties of a perovskite material (CH$_{3}$NH$_{3}$PbI$_{3}$). Using non-equilibrium Green functions formalism, we calculate the local density-of-states for several values of the electron-phonon scattering strength as well as in the case of ballistic transport. We report an Urbach-like penetration of the density-of-states in the bandgap due to scattering. The density-of-states expands deeper in the bandgap with the scattering strength. We determine the electronic current contributions relative to photon absorption and photon emission. This allows to estimate the Urbach energy from the absorption coefficient below the bandgap. Values of Urbach energy up to 9.5 meV are obtained, meaning that scattering contribution to the total experimental Urbach energy of 15 meV is quite important. The Urbach tail is generally attributed to the disorder of the system, leaving hope for further improvement. Yet, we show in this work that a significant contribution to the Urbach tail comes from carriers-phonon interactions, an intrinsic properties of the material which thus set a lower bound on the improvement margin. Finally, we estimate the open-circuit voltage $V_{oc}$ for a solar cell assuming such a material as an absorber. $V_{oc}$ losses increase with the scattering strength, up to 41 mV. This study then confirms that scattering of electrons in perovskite material modifies the optical behaviour.

\end{abstract}

\section{Introduction}

Perovskite materials are widely studied for both device conception and theoretical understanding. Among the variety of perovskites materials, organic-inorganic hybrid perovskites are the most promising materials for optoelectronic applications. First photovoltaic cells made of hybrid perovskites lacked stability and had low solar conversion efficiency, of order 3\%. Nowadays, fabrication techniques of perovskites have been largely improved, making them more stable and reliable to use in solar cells. Production costs have also been drastically reduced, paving the way to further decrease of the PV electricity price \cite{Prix}. This fast evolution of technology leads to the increase of efficiency, up to 24\% in less than 10 years \cite{Elumalai}, and have reached over 28\% in Si-perovskite tandem devices \citep{PV28}. 
\\\indent
Hybrid perovskite represents one of the most promising materials for reliable solar conversion technology thanks to their excellent optical and electrical properties. They exhibit high carrier diffusion lengths, reaching up to 1 $\mu m$ even in nanostructures \cite{Diff}. High carrier lifetime has also been measured, to be around several $\mu s$ \cite{Time}. Moreover, it has been shown that hybrid perovskites have remarkable photon absorption properties \cite{Abs}. These properties have also been observed in perovskite layers as thin as 50 nm \cite{Abs3}. The main advantage of such thin films is to be integrated in light-weighted, flexible and transparent solar cells \cite{Abs2} that could find applications in wearable electronics.
\\\indent However, the remarkable properties of these materials cannot be fully understood considering the peculiar structure of hybrid-perovskites. On the one hand, with these excellent carrier transport properties, we could expect hybrid-perovskites to exhibit high carrier mobilities. Yet, several groups have reported low values for carrier mobilities in perovskite-based devices. Y. Mei \textit{et al} found mobilities of $1$ cm$^{2}$/V.s in a field effect transistor \cite{Mob1}. D. Shi \textit{et al} estimated the electron mobility in a single crystal to be $115$ cm$^{2}$/V.s \cite{Mob2}. Microwave conductivity measurements gave carrier mobilities from $6$ to $60$ cm$^{2}$/V.s \cite{Mob3}\cite{Mob4}. These measured values are quite low compared to mobilities in other inorganic materials used in PV cells (electron mobility is equal to $1500$ cm$^{2}$/V.s in silicium), and could indicate some disorder in the crystalline structure.
On the other hand, it is possible to extract information about the quality of the material from the absorption features below the bandgap. The Urbach rule gives the absorption coefficient below the bandgap as a function of photon energy and temperature. It is described by a "tail" with the expression :
\begin{equation}
\alpha(h\nu) = \alpha_{0} \, \mathrm{exp}\left[\frac{h\nu - E_{g}}{E_{U}}\right]
\end{equation}
where $\alpha_{0}$ is a constant, $E_{g}$ is the material bandgap, $h\nu$ is the incident photon energy, and $E_{U}$ is called the Urbach energy, corresponding to the width of the exponential part of the absorption below the bandgap. This empirical formula is an important tool to study the optical absorption, due to its simplicity and its applicability for various types of materials and structures. However, the origin of this phenomenon is still debated in the literature. It could be caused by several types of disordering that modify the shape of the density-of-states (DOS) band edge \cite{tails}. Presence of impurities in the material can induce localized states near the band edges as well \cite{Impurity}. Electron-phonon interactions have also been associated with Urbach effect \cite{Urb1,grein_temperature_1989}. Recently, using an approach very close to ours, all such behaviors have been considered allowing to reproduce the band tail and the band gap narrowing in nanowire \cite{sarangapani_non-equilibrium_2019}. The tail, particularly, can provide informations about the quality of the nanostructure. In well-ordered materials like GaAs, the Urbach energy is as low as 7 meV \cite{Urb2}, while in amorphous materials, it can be much higher, up to 50 meV \cite{Urb3}. In the case of hybrid perovskite-based device, the Urbach energy has been estimated experimentally to 15 meV \cite{EUP}, confirming the well-ordered structure of such a material. The so called Urbach effect modifies the characteristic of the corresponding solar cell with a Stokes shift and an impact on the open-circuit voltage $V_{oc}$ \cite{WCPEC}. The definition of the optical bandgap can also be discussed \cite{Rau}, \cite{Perovgap}.
\\\indent 
Low values of mobilities as well as Urbach-like behaviours can be related to electron-phonon scattering (EPS). In hybrid perovskites, the scattering with polar optical phonons is expected to be large since the difference between $\epsilon_{\infty}$ and $\epsilon_{stat}$ (the dynamic and the static dielectric constant respectively) is very large \cite{Dielectric} (values reported in Table I). Moreover, rotations of the CH$_{3}$NH$_{3}$ organic matrix may play a key role in quantum transport properties \cite{Rot}. Random rotations of cations induce fluctuations of the electrostatic potential. Electrons and holes in the material becomes localized and spatially separated, affecting their transport properties. The understanding of these mechanisms is necessary for a controlled engineering of perovskite-based structures, and thus thus lead to further improvements of devices' efficiencies.

In this work, we propose a numerical study of the EPS in hybrid-perovskite materials. Using non-equilibrium Green's functions formalism \cite{Green,sarangapani_non-equilibrium_2019}, we show that such an interaction modifies the electronic DOS, leading to the enhancement of an Urbach tail in the bandgap. We also develop an analytic model of the DOS based on Green's functions. This model, which is in good agreement with the numerical results, highlights the impact of different parameters such as effective masses and phonon energy on the Urbach behaviour of the material. Calculations of the absorption and the emission current contributions finally allow us to estimate the variation of $V_{oc}$ with respect to the EPS strength.

\begin{table}[b]
\caption{Constant parameters of CH$_{3}$NH$_{3}$PbI$_{3}$}
\centering
\begin{tabular}{ll}
\hline
Parameter & Value \\
\hline
m$^{*}_{e}$/m$_{0}$ & $0.23$ \\
m$^{*}_{h}$/m$_{0}$ & $0.17$ \\
$\epsilon_{\infty}$ & $6.0$ \\
$\epsilon_{stat}$ & $29.0$ \\
E$_{pop}$ (meV) & $8.0$ \\
\hline
\end{tabular}
\end{table}

\section{Methodology}
In order to study the impact of EPS, we made numerical calculations using non-equilibrium Green's function formalism within the self-consistent Born approximation \cite{cavassilas_one-shot_2013}. This model is detailed in ref \cite{Green}, but we recall here the main elements needed for an easier understanding. We consider a three-dimensional (3D) device, with wide transverse dimensions perpendicular to the transport axis. This allows to assume a one-dimensional (1D) model in the longitudinal direction for each transverse mode corresponding to a propagation in the transverse plane with a given transverse wave vector. In our case, we consider N$_{k_{t}}$ = 30 transverse modes.  For each mode, we define a Hamiltonian H$_{k_{t}}$, in the 1D site basis using the effective mass approximation which which is suitable for the study of most crystalline material, including perovskite \cite{even_importance_2013}. This allows to calculate the corresponding lesser G$^{<}$(E, ${k_{t}}$) and greater G$^{>}$(E, $k_{t}$) Green's functions, relative to electron and hole densities respectively. The DOS in both conduction and valence bands is given by the retarded Green function G$^{R}(E,k_{t})$. We consider the diagonal approximation to describe the EPS and assume the effective mass approximation for both conduction and heavy-hole valence bands. In order to make the calculations with such formalism, we need to take the band diagram of the device as well as the EPS strength and phonon density as input data in our code. For calculations in a solar cell, the band diagram, relative to electric field and band-offset is obtained by a Poisson-Schr\"odinger self-consistent approach. To do so, we have implemented the structure in SCAPS software \cite{SCAPS}. SCAPS is a solar cell simulation program which calculates optical and electronic characteristics of multi-layer semiconductor devices based on a drift-diffusion model. In our case, we have used SCAPS to extract data relative to the band diagram of our device as an input file of our Green's functions simulations.
\\\indent
In such a polar material, polar optical phonons dominate carrier transport. While the model of Sarangapani \textit{et al} includes scatterings with both polar optical phonon and charged impurities, and Grein \textit{et al} assumed only acoustic phonons, we then only assume the polar optical phonons. The strength of this electron-phonon interaction is given by :
\begin{equation}
M_{pop} = C_{EPS} \frac{1}{2\pi S}\frac{E_{pop}}{4\pi \epsilon_{0}}\left(\frac{1}{\epsilon_{\infty}} - \frac{1}{\epsilon_{stat}}\right)
\end{equation}
where S is the surface of the transverse cross-section of radius R = 30nm, $E_{pop}$ is the polar optical phonon energy of the material, $\epsilon_{0}$ is the vacuum permittivity, $\epsilon_{\infty}$ and $\epsilon_{stat}$ represent the relative dynamic and static permittivity respectively. All used parameters are given in Table I. C$_{EPS}$ is a numerical fitting factor, adjusted to a value such that the model accounts for the correct value of the electron mobility \cite{Niquet}. The relation between C$_{EPS}$ and the mobility is shown in Table II. Electron mobility in hybrid perovskites has been experimentally measured from 1 to 100 cm$^{2}$/V.s. Mechanisms controlling the carrier mobility in this material are not clearly understood, since the CH$_{3}$NH$_{3}$ molecular rotations could have an impact on carrier transport properties \cite{Rot}. From a formalism point of view, these rotations are close to the electron-phonon interaction, but with unknown parameters. In the following, we perform calculations with 4 different values of C$_{EPS}$ corresponding to mobilities ranging from 0.13 to 86 cm$^2$/V.s as reported in the literature.
\\\indent
The number of phonons per unit cell $\mathcal{N}_{pop}$ is given by the Bose-Einstein distribution :
\begin{equation}
\mathcal{N}(E,T) = \frac{1}{\mathrm{exp}\left(\frac{E}{k_{B}T}\right)-1}
\end{equation}
with $E=E_{pop}=8$ meV and $T=300$ K in the case of phonons in CH$_{3}$NH$_{3}$PbI$_{3}$. Since $E_{pop}$ is small in the considered perovskite material (compared to $k_{B}T=25.7$ meV and compared for example to the GaAs value of $35$ meV), $\mathcal{N}_{pop}$ is large. We consider the EPS by calculating the corresponding interaction self-energies :
\begin{equation}
\Sigma_{pop}^{<}(E,k_{t}, E_{pop}) = \sum_{k_{t}'} M_{t}(k_{t},k_{t}')
[ \mathcal{N}_{pop} G^{<}(E-E_{pop}, k_{t}') + (\mathcal{N}_{pop}+1) G^{<}(E+E_{pop}, k_{t}') ] M_{t}(k_{t},k_{t}')
\end{equation}
\begin{equation}
\Sigma_{pop}^{>}(E,k_{t}, E_{pop}) = \sum_{k_{t}'} M_{t}(k_{t},k_{t}') [ \mathcal{N}_{pop}
G^{>}(E+E_{pop}, k_{t}') + (\mathcal{N}_{pop}+1) G^{>}(E-E_{pop}, k_{t}') ] M_{t}(k_{t},k_{t}')
\end{equation}
where $M_{t}(k_{t},k_{t}')$ is the intermode coupling matrix, which depends on both considered modes $k_{t}$ and $k_{t}'$, and on M$_{pop}$ \cite{Manel}. Equ.4 describes the interaction with electrons while Equ.5 accounts for those with holes. Terms in $\mathcal{N}_{pop}$ correspond to phonon absorption while terms in ($\mathcal{N}_{pop}$+1) correspond to phonon emission. Calculations of self-energies are self-consistent, since G$^{<}$ and G$^{>}$ also depends on these self-energies \cite{Green}.

\begin{table}[b]
\caption{Numerical values of the scattering strength multiplicative coefficient, with the estimated electron mobility and calculated Urbach energy}
\centering
\begin{tabular}{lll}
\hline
C$_{EPS}$ & $\mu e^{-}$  ($cm^{2}/V.s$) & $E_{U} (meV)$ \\
\hline
$1.6$ & $86$ & $5.25$ \\
$5.0$ & $18$ & $6.6$ \\
$10.0$ & $1.7$ & $7.9$ \\
$20.0$ & $0.13$ & $9.5$ \\
\hline
\end{tabular}
\end{table} 
 
NEGF calculations allows us to estimate the current generated by the cell in the dark as well as under illumination. From currents calculations, we can thus deduce the relation between V$_{oc}$ and the EPS.

The electrical characteristic of a solar cell is close to the diode characteristic. 
In the dark, the current generated by the cell is given by :
\begin{equation}
\mathcal{I}_{dark} = I_{0} \, \left[ \mathrm{exp}\left( \frac{qV}{k_{B}T}\right) -1 \right]
\end{equation}
with $I_{0}$ the current relative to recombination of majority carriers in the cell for a bias $V$=0V. In the dark, $I_{0}$ compensates the minority carrier current, and the total current is zero. Under illumination, the current in the cell at a bias $V$ can be estimated by :
\begin{equation}
\mathcal{I}_{light} = -I_{sc} + \mathcal{I}_{dark}
\end{equation}
where $I_{sc}$ is the short-circuit current of the cell, which is the photocurrent at zero bias. Here, we consider that the photocurrent does not depend on the bias $V$. It is then possible to estimate the open-circuit voltage $V_{oc}$ only with the Green's functions calculated at $V$=0V :
\begin{equation}
V_{oc}= k_{B}T \, \mathrm{ln}\left( \frac{I_{sc}}{I_{0}} \right).
\end{equation}
$V_{oc}$ is calculated for each value of C$_{EPS}$, and each $V_{oc}$ is compared to the ballistic counterpart (C$_{EPS}$=0) to estimate $\delta V_{oc}$, the losses due to EPS. 
$I_{sc}$ and $I_{0}$ are calculated using :
\begin{equation}
I_{sc} = \int_{h\nu} I_{sc}(h\nu) \mathrm dh\nu
= \sum_{k_{t}} \int_E \int_{h\nu}  q \, \Omega \, \mathcal{C}_{c-v} \, \rho(h\nu) \, G^{<}(E,k_{t})
G^{>}(E+h\nu,k_{t}) \, \mathcal{N}(h\nu,T_{sun}) \, dE \,  dh\nu ,
\end{equation}
and
\begin{equation}
I_{0} = \sum_{k_{t}} \int_E \int_{h\nu} q \, \Omega \, \mathcal{C}_{c-v} \, \rho(h\nu) \, G^{<}(E+h\nu,k_{t})
 G^{>}(E,k_{t}) \, dE \, dh\nu ,
\end{equation}
where $\mathcal{C}_{c-v}(h\nu)$ is a band coupling parameter\cite{Rosencher}, $E$ is the electron energy, $h\nu$ is the photon energy and $\Omega$ is the solid angle of the light source, which is equal to $6.79 \times 10^{-5}$sr for $I_{sc}$ and $\pi$ sr for $I_{0}$. $\mathcal{N}(h\nu,T_{sun})$ is the Bose-Einstein distribution of photons (see Equ. 3), at the energy $h\nu$ and temperature $T_{sun}$=6000K, and $\rho(h\nu)$ represents the photon DOS per unit energy and unit volume :

\begin{equation}
\rho(h\nu) = \frac{8 \pi }{(hc)^{3}} (h\nu)^{2}.
\end{equation}

$I_{0}$ depends on $G^{<}(E+h\nu)$ and $G^{>}(E)$, which are the electron density at energy $E+h\nu$ and the hole density at energy $E$ respectively. An electron in the conduction band can recombine with a hole in the valence band, emitting a photon of energy $h\nu$. On the other hand, Equ. 9 shows that $I_{sc}$ depends on $G^{<}(E)$ and $G^{>}(E+h\nu)$, which are the electron density at energy $E$ and the hole density are energy $E+h\nu$. An electron in the valence band can interact with a photon of energy $h\nu$ to populate a state in the conduction band.

\section{Results}

\subsection{Density-of-states}

Figure 1 shows the local-DOS calculated in a solar cell with an hybrid perovskite active region. We performed the calculations with a C$_{EPS}$ corresponding to an electron mobility of 86 cm$^{2}$/V.s and in the ballistic approximation (C$_{EPS}$=0). The two situations are compared in Figure 1. In the ballistic case, quantum tunneling leads to a small but finite penetration of the DOS in the bandgap. This DOS distribution has been shown to impact optical properties of ultra thin solar cells \cite{Tunnel}. By comparison, the presence of EPS strongly enforces the penetration of DOS within the bandgap, which largely dominates the one induced by quantum tunneling. The presence of these near-band edge states allow both electrons and holes to be either photogenerated or recombined at energies lower than the material bandgap of 1.55eV. In order to illustrate the evolution of broadening with the interaction strength, we have also made local-DOS calculations for each value of the scattering strength. Figure 2 shows these DOS at the middle of the device \textit{versus} the electron energy. The DOS tends to penetrate deeper in the bandgap as the EPS becomes stronger. 
\begin{figure}[!htb]
        \center{\includegraphics[scale=1.2]
        {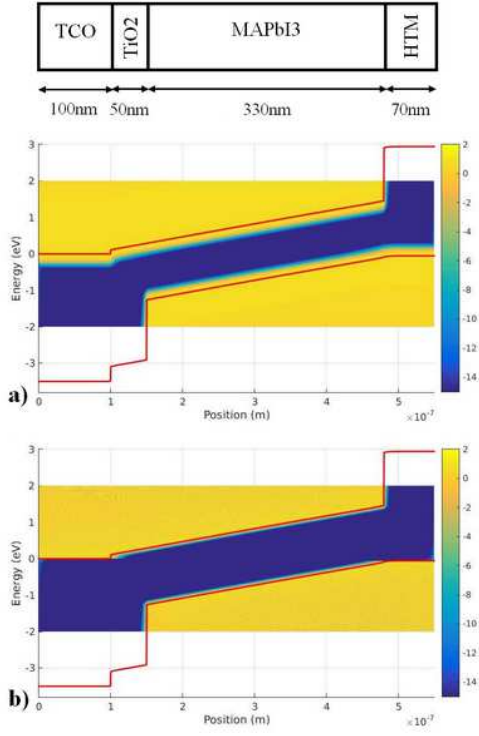}}
        \caption{\label{fig1} Local density-of-states in a perovskite-based device (shown above) as a function of position and electron energy, calculated a) with electron-phonon interaction, b) without electron-phonon interaction.}
      \end{figure}
%      \begin{figure}[!htb]
%        \center{\includegraphics[scale=0.35]
%        {figures/Figure1.png}}
%        \caption{\label{fig2} Local density-of-states (LDOS) in a perovskite-based device as a function of position and electron energy, calculated without electron-phonon interaction.}
%      \end{figure}

In order to get some physical insight, we developed an analytic model showing the relative importance of the various parameters. The gap density of states is related to the imaginary part of the retarded Green's function :
\begin{equation}
\mathrm{D_{k_{t}}}(E) = \frac{1}{2\pi} \, \mathrm{Im}(G^{R}) = \frac{\Gamma_{k_{t}}}{(E-E_{k_{t}})^{2}+\Gamma_{k_{t}}^{2}}
\end{equation}
where $E_{k_{t}}$ is the energy of the band edge (we consider here the conduction band) and $\Gamma_{k} = \frac{1}{2\pi} \, \mathrm{Im}(\Sigma^{R})$. Using Equs.4 and 5 and the approach developed in Ref.\cite{Manel}, we can write Equ.2 in a continuous form :
\begin{equation}
\Gamma_{k_{t}} = \frac{M_{t}}{(2\pi)^{3}}\int \frac{d^{3}k_{t}'}{|k_{t}-k_{t}'|^{2}}\left[\mathcal{N}_{pop} \, \mathrm{D_{k_{t}'}}(E+E_{pop})+(\mathcal{N}_{pop}+1) \, \mathrm{D_{k_{t}'}}(E-E_{pop}) \right]
\end{equation}
which is valid for small electron concentrations. As shown in the Supplementary Material (Equs.S5 and S9) this set of equations can be reduced to :
\begin{equation}
\mathrm{D}_{0}(E) = \frac{\Gamma_{0}(E)}{E^{2}} = \left( \frac{E_{M}}{E}\right)^{\frac{3}{2}}\left[\mathcal{N}_{pop} \, \mathrm{D}_{0}(E+E_{pop})+(\mathcal{N}_{pop}+1) \, \mathrm{D}_{0}(E-E_{pop}) \right]
\end{equation}
with $E^{\frac{3}{2}}_{M_{t}} = 0.02 \, M^{2} \, \sqrt{\left(\frac{2m^{*}}{\hbar^{2}}\right)}$.

This is a set of ladder equations which leads to an exponential decrease at large negative energies. In this regime, $|E|>>E_{pop}$ and Equ.14 becomes a simple differential equation, valid at large $\frac{|E|}{E_{pop}}$. To solve it, we use the expected exponential behaviour by writing $\mathrm{D}(E)=A \, \mathrm{exp}(\sigma(E))$ and inject it into Equ.14 together with a first order expansion $\sigma(E\pm E_{pop}) = \sigma(E) \pm \sigma'(E) E_{pop}$ . With this, all terms in $\sigma(E)$ disappear and, by expanding $\mathrm{exp}(\pm \sigma'(E) E_{pop})$ to second order and keeping the dominant term in $|E|$, one gets :
\begin{equation}
E_{pop} \, \sigma'(E) = \sqrt{\left(\frac{2}{2 \, \mathcal{N}_{pop}+1}\right)} \left( \frac{|E|}{E_{M}} \right)^{\frac{3}{4}}
\end{equation}
whose integration gives :
\begin{equation}
\mathrm{ln}(\mathrm{D}_{0}) = \sigma = - \sqrt{\left(\frac{2}{2 \, \mathcal{N}_{pop}+1}\right)} \, \frac{4}{7} \, \frac{(E-E_{g})^{\frac{7}{4}}}{E_{pop}E^{\frac{3}{4}}_{M}} + \mathrm{C}te .
\end{equation}

This depends upon two basic parameters $E_{pop}$ and $E_{M}$. Note that it depends upon $1/M_{t}$ linearly.  This analytic expression is compared to the full numerical result in Figures 3 and 4, showing an excellent agreement at large energies, apart from a shift due to the constant. In particular, we confirm that the slope of the Urbach tail is proportional to $E_{pop}$ and also to $E_{M}^{\frac{3}{4}}$ which is linear in $M_{t}$ and that one can relate to the mobility.
\\\indent
As seen before, DOS tends to penetrate deeper in the bandgap as the scattering with optical phonons becomes more important. Another interesting result shown Figure 4 is that the polar optical phonon energy also plays an important role. The higher is the phonon energy, the deeper the DOS expands below the band-edge. Since $E_{pop}$ in perovskites is low, DOS penetration in the bandgap is finally quite modest.
      \begin{figure}[!hb]
        \center{\includegraphics[scale=1.0]
        {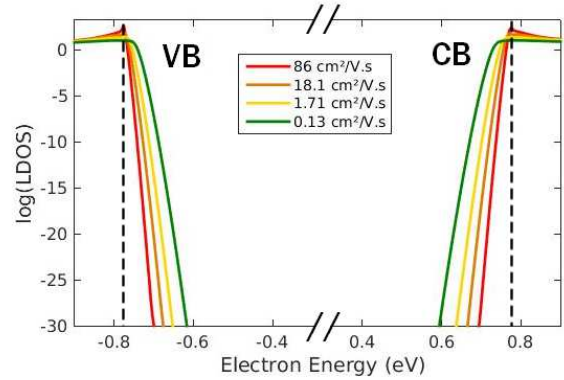}}
        \caption{\label{fig3} Logarithm of the density-of-states as a function of electron energy at a fixed position in the middle of the active region of the perovskite-based device. Each color corresponds to a different scattering coupling strength, and thus to a different value of electron mobility}
      \end{figure}
\begin{figure}[!hb]
        \center{\includegraphics[scale=1.0]
        {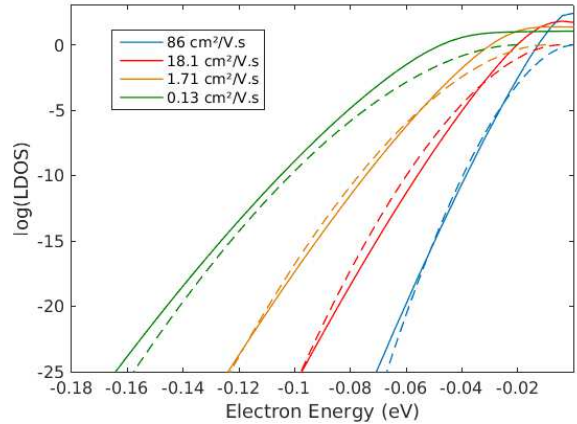}}
        \caption{\label{fig4} Density-of-states in the considered device as a function of the electron energy. Each color corresponds to a value of the EPS strength (and therefore of the electron mobility). Plain lines correspond to the self-consistent calculations. Dashed lines correspond to the calculated DOS using Equ. 16. The conduction band edge has been shifted in energy at 0 eV.}
\end{figure}
\begin{figure}[!hb]
        \center{\includegraphics[scale=1.0]
        {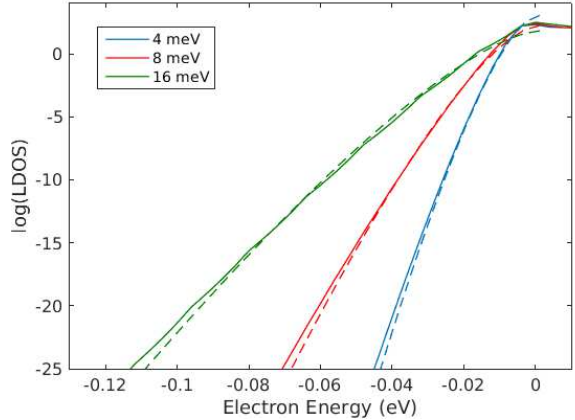}}
        \caption{\label{fig5} Density-of-states in the considered device as a function of the electron energy. Each color corresponds to a value of the optical phonon energy. Plain lines correspond to the self-consistent calculations. Dashed lines correspond to the calculated DOS using Equ. 16. The conduction band edge have been shifted in energy at 0 eV.}
\end{figure}
\\\indent
\subsection{Urbach Energy}

Using our results, we can now estimate the Urbach energies by calculating the exponential decay of the absorption coefficient $\alpha$. Starting from the calculation of the photogenerated current $I_{sc}$ (Equ.9), we obtain $\alpha$ expressed as a function of the photon energy $h\nu$ :
\begin{equation}
\alpha (h\nu) = - \frac{1}{a} \, \mathrm{ln} \left( 1- \frac{a \, I_{sc}(h\nu)}{q \mathcal{N}_{h\nu} \, \rho(h\nu) \, L \, v_{e}(h\nu) \, dh\nu}  \right)
\end{equation}
where $L$ is the thickness of the device, $v_{e}(h\nu)$ is the group velocity of electrons generated by a photon of energy $h\nu$, $a$ and $\mathrm dh\nu$ correspond to numerical step in position and energy respectively. Figure 5 shows $\alpha$ for each C$_{EPS}$ as a function of the photon energy. It confirms the Urbach-like exponential decrease of the DOS inside the bandgap, which is all the more severe as EPS is large. The corresponding Urbach energies are reported in Table II. Despite the very low $E_{pop}$, the calculated $E_{U}$ are not negligible, up to 9.5 meV, compared to the experimental value of 15 meV. Yet, this value is generally explained by crystalline disorder ; our result shows that EPS also impact the Urbach behaviour. However, contrary to disorder, scattering with phonons is intrinsic, and therefore unavoidable. 

% The corresponding Urbach energies are given in Table 2. Urbach energies are lower than the experimental values of 15 meV, even with a strong coupling. This is interesting since hybrid perovskite being highly polar material (cf Table 1), we could expect the interaction between electrons and polar optical phonons to be important. However, the phonon energy is only 8mV, therefore as shown in the previous chapter, effect of scattering on the Urbach behaviour is low. Our results show that EPS is not the main cause of Urbach tails in hybrid perovskite.

%\begin{table}
%\caption{Numerical value of the scattering strength multiplicative coefficient, with associated electron mobility, calculated Urbach energy and $V_{oc}$ losses (compared to the ballistic counterpart).}
%\centering
%\begin{tabular}{lll}
%\cmidrule(r){1-4}
%C$_{EPS}$ & $µe^{-}$  ($cm^{2}/V.s$) & $E_{U} (mV)$ \\
%\midrule
%$1.6$ & $86$ & $5.25$ \\
%$5.0$ & $18$ & $6.6$ \\
%$10.0$ & $1.7$ & $7.9$ \\
%$20.0$ & $0.13$ & $9.5$ \\
%\bottomrule
%\end{tabular}
%\end{table}

%            \begin{figure}[!htb]
%        \center{\includegraphics[scale=0.35]
%        {figures/Figure3.png}}
%        \caption{\label{fig:my-label} Logarithm of the absorption spectra calculated for each value of the electron-phonon scattering strength. Dashed line correspond to the bandgap of CH$_{3}$NH$_{3}$PbI$_{3}$ at 1.55 eV}
%      \end{figure}
      
      \begin{figure}[!htb]
        \center{\includegraphics[scale=1.0]
        {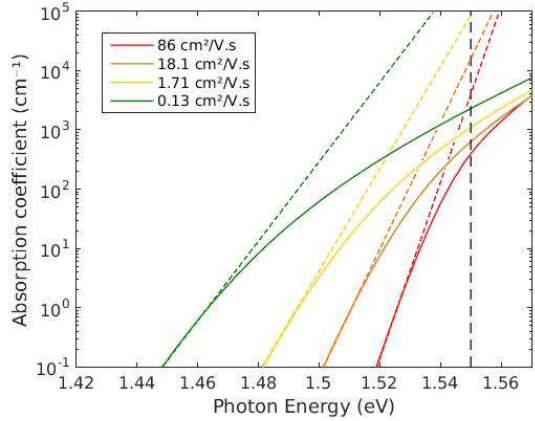}}
        \caption{\label{fig6} Absorption coefficient $\alpha$ (solid lines) versus photon energy calculated for each value of the electron-phonon scattering strength. Colored dashed lines correspond to the slopes used to calculate Urbach energies. Black dashed line correspond to the bandgap of CH$_{3}$NH$_{3}$PbI$_{3}$ at 1.55 eV}.
\end{figure}
      
%      \begin{table}
%\caption{Calculated values of the open-circuit voltage V$_{oc}$ and V$_{oc}$ losses for each considered electron mobility}
%\centering
%\begin{tabular}{llr}
%\cmidrule(r){1-3}
%$µe^{-}$  ($cm^{2}/V.s$) & $V_{oc} (V)$ & $\delta V_{oc} (mV)$ \\
%\midrule
%ballistic & $1.0477$ & $0$ \\
%$86$ & $1.0466$ & $-1.1$ \\
%$18$ & $1.045$ & $-2.7$ \\
%$1.7$ & $1.0348$ & $-12.9$ \\
%$0.13$ & $1.0068$ & $-40.9$ \\
%\bottomrule
%\end{tabular}
%\end{table}

            \begin{figure}[!htb]
        \center{\includegraphics[scale=1.0]
        {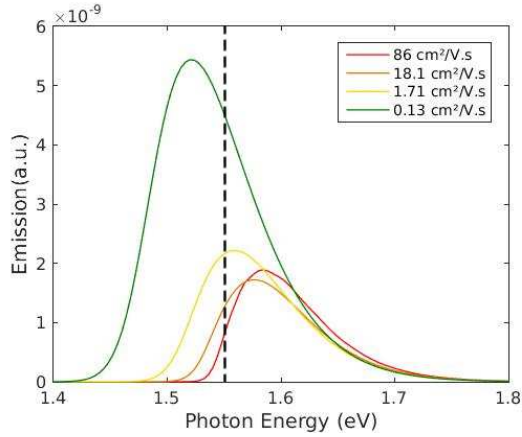}}
        \caption{\label{fig7} Emission spectra calculated for each value of the electron-phonon scattering strength.}
      \end{figure}

\subsection{Emission and $V_{oc}$}

Figure 6 shows the calculated emission spectra for each value of the scattering strength, calculated with Equ. 10. We report both a red-shift and an increase of the intensity with the EPS strength, . This is another consequence of the DOS penetration. Electrons in conduction band can occupy the states below the bottom of conduction band, to recombine with holes above the top of valence band, emitting photons of lower energies than the material bandgap. In an efficient photovoltaic device, we want to absorb as many photons as possible while preventing as much as possible electron-hole recombinations. In order to quantify the impact of scattering on the open-circuit voltage, we calculate the ratio of absorption over emission (needed in Equ. 8). Figure 7a shows the calculation of this ratio as a function of C$_{EPS}$. The ratio decreases for higher values of the phonon coupling. The penetration of the DOS due to EPS therefore favors the recombination and thus the photon emission. This confirms that the degradation of the sharpness of the absorption coefficient due to Urbach tail tends to deteriorate the quality of the material for photovoltaic applications \cite{Rau}. Decrease of the absorption over emission ratio with the EPS can be linked to Urbach behaviour. Corresponding values of $\delta V_{oc}$ are also shown in Figure 7b. We then confirm that $V_{oc}$ decreases with the EPS up to 41 mV for the highest C$_{EPS}$, therefore for the lowest mobility. Fortunately, this degradation is limited by the low value of $E_{pop}$ in the studied perovskite material. However, this value is not negligible and such a behaviour is intrinsic and generalizable to all semiconductors.

%      \begin{figure}[!htb]
%        \center{\includegraphics[scale=0.35]
%        {figures/Figure5.png}}
%        \caption{\label{fig8} Absorption/Emission ratio for each value of the electron mobility considered.}
%      \end{figure}
            \begin{figure}[!htb]
        \center{\includegraphics[scale=0.7]
        {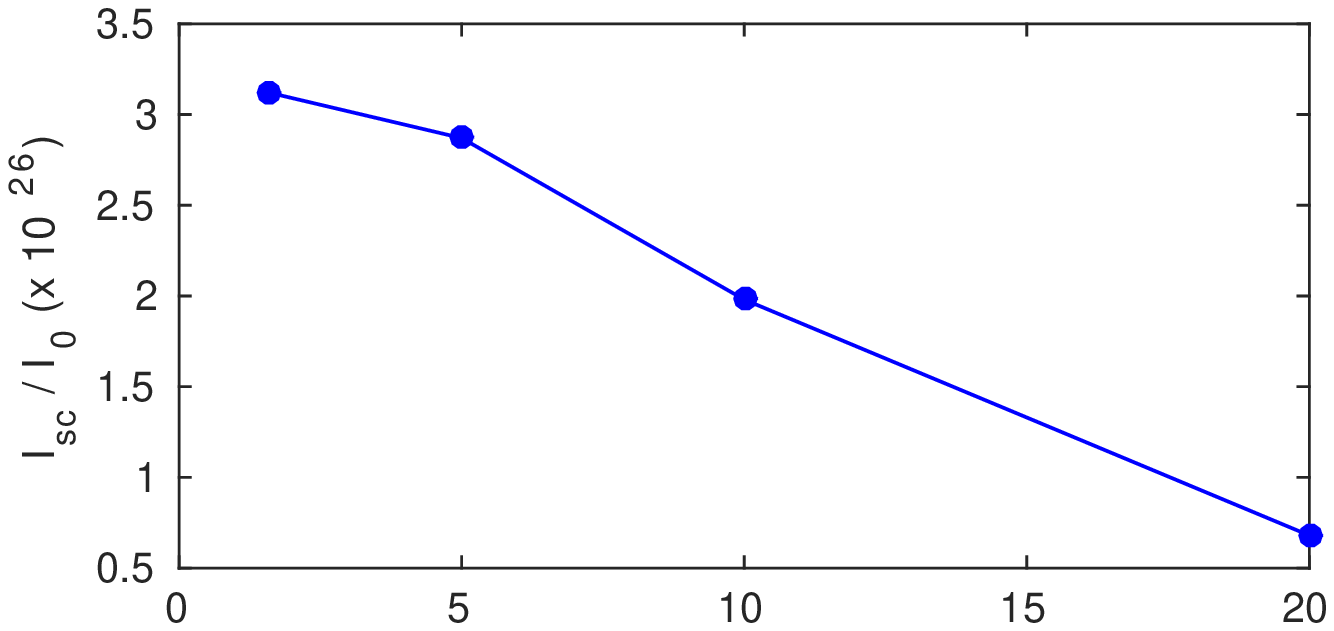}
        		\includegraphics[scale=0.7]
        {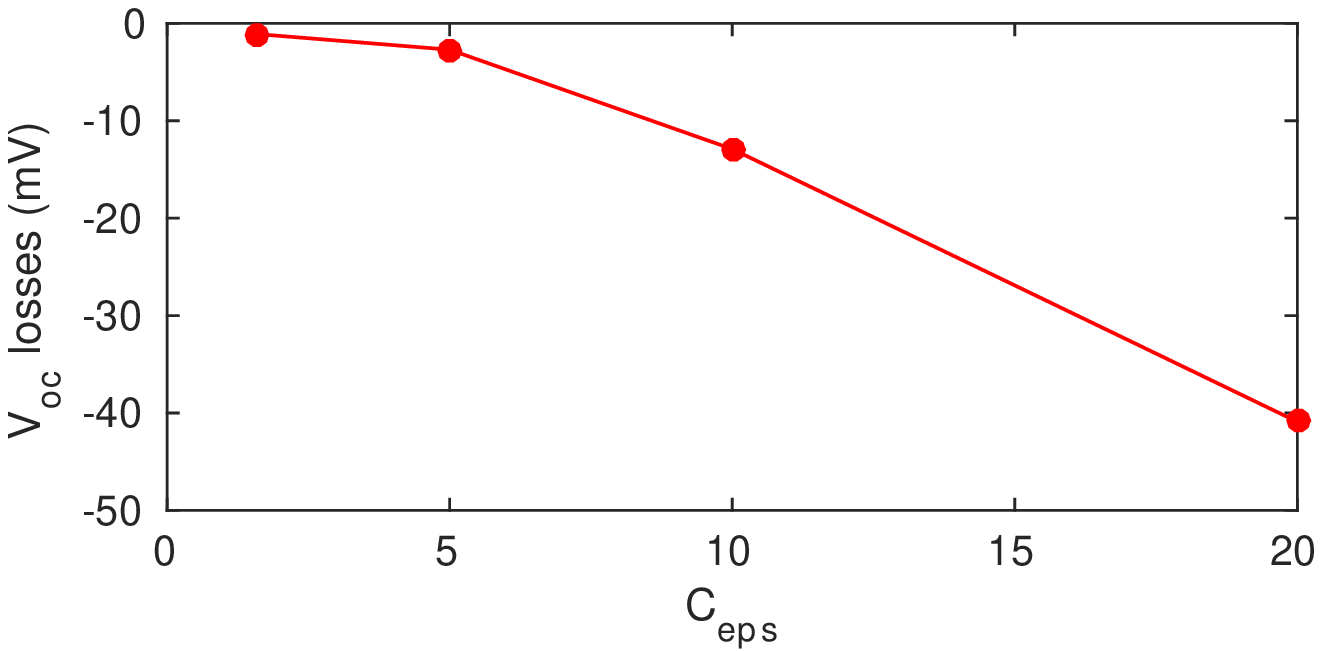}}
        \caption{\label{fig8} Top : $I_{sc}/I_{0}$ ratio for each value of $\mathrm{C_{EPS}}$ considered. Bottom : $V_{oc}$ losses due to electron-phonon scattering for each value of $\mathrm{C_{EPS}}$ considered.}
      \end{figure}

\section{Conclusion}
In this paper, we studied the impact of the electron-phonon scattering on optical and electrical properties of an hybrid perovskite. Our numerical calculations showed that electron-phonon scattering induces a penetration of the density-of-states in the bandgap, which increases with the scattering strength. Such modifications induce an energy shift of the emission spectra, since photogenerated electrons and holes can recombine at lower energies than the material bandgap. We have also investigated the Urbach effect induced by electron-phonon scattering. We obtained values of Urbach energies from 5.25 to 9.5eV, which means that phonon scattering contributes significantly to the exponential decrease of absorption below the bandgap. The analytic description of the density of states confirms this strong impact on the Urbach behaviour since it depends on both the scattering strength and the optical phonon energy. Finally, we calculated the open-circuit voltage $V_{oc}$ for each scattering strength compared to the ballistic counterpart. We obtained $V_{oc}$ losses up to 41 mV due to the electron-phonon scattering. In order to tweak the results concerning the impact of electron-lattice interactions on the optical and electrical properties, a more precise inclusion of CH$_3$NH$_3$ rotations should be led. Moreover, the study of the influence of temperature can also provide relevant information about the carrier transport properties in such a material.

\section{Supporting Information}
Calculation details allowing to obtain the Equ. (14) of the article.

We want to get a simplified version of Equ.13. For this, we first integrate it over angles, getting for each term :
\begin{equation}
J(E) = \int \frac{d^{3}k'}{|k'-k|^{2}} \mathrm{D}_{k'}(E) = 2 \pi \int_{0}^{\infty} \mathrm{D}_{k'}(E) \frac{k'}{k} \mathrm{ln} \left( \frac{k+k'}{|k-k'|}\right) dk'
\end{equation}
The integrand exhibits a logarithmic singularity which is not important for integral properties. To avoid the problem, we integrate Equ.S1 by parts and obtain :
\begin{equation}
J(E) = 2 \pi \int_{0}^{\infty} \frac{d\mathrm{D}_{k'}(E)}{dk'} \mathrm{F}(k,k') dk'
\end{equation}
with
\begin{equation}
\mathrm{F}(k,k') = k \left\{ \frac{1}{2} \left[ \left( \frac{k'}{k} \right)^{2} -1 \right] \mathrm{ln} \left( \frac{k+k'}{|k-k'|} \right) + \frac{k_{0}}{k} \right\}
\end{equation}
where the integrated term $\mathrm{D}_{k'} \mathrm{F}(k,k')$ vanished since $\mathrm{F}(k,k')$ tends to 0 when $k'=0$ and $\mathrm{D}_{k'}$ tends to 0 when $k'$ tends to $\infty$. Now, we use the fact that $\frac{d\mathrm{D}_{k'}}{dk'}$ presents a strong peak at $k'=k_{0}$ so that one can write :
\begin{equation}
J(E) = 2 \pi \mathrm{F}(k,k_{0}) \mathrm{D}_{0}(E)
\end{equation}
where $k_{0}(E)$ is some function of E to be determined in the following. To inject Equ.S4 in Equ.13 we have to determine $k_{0}(E+E_{pop})$ but $E_{pop}$ is small and Equ.3 is almost symmetric in $E+E_{pop}$ so that we take a common value $k_{0}(E)$ for the two terms in Equ.13. This gives :
\begin{equation}
\Gamma_{k} = \frac{M^{2}}{4 \pi ^{2}}\left\{ \mathrm{N}_{pop} \mathrm{D}_{0}(E+E_{pop})+ (\mathrm{N}_{pop}+1) \mathrm{D}_{0} (E-E_{pop}) \right\} \mathrm{F}(k,k_{0})
\end{equation}
We now want to evaluate $k_{0}(E)$ by determining the peak of $\frac{d\mathrm{D}_{k'}}{dk'}$. We thus need the general expression of $\mathrm{D}_{k'}(E)$ which one can write :
\begin{equation}
\mathrm{D}_{k'}(E) \simeq \frac{\Gamma_{k'}(E)}{(|E-E_{k'}|)^{2}}
\end{equation}
valid for large negative E. A full derivation requires the knowledge of $\Gamma_{k'}$. From Equ.S5, we know that $\Gamma_{k_{0}} = \frac{\Gamma_{0}}{2}$ since $\mathrm{F}(k_{0},k_{0})= \frac{1}{2}\mathrm{F}(0,k_{0})$. To account for this, we write Equ.S6 :
\begin{equation}
\mathrm{D}_{k'}(E) \simeq \frac{\Gamma_{0}\left(1-\frac{k'^{2}}{2k_{0}^{2}}\right)}{(|E|+E_{k'})^{2}} = \mathrm{D}_{0}(E)\frac{\left(1-\frac{k'^{2}}{2k_{0}^{2}}\right)}{\left(1+\frac{k'^{2}}{k_{E^{2}}}\right)^{2}}
\end{equation}
We now search for the peak in $\frac{d\mathrm{D}_{k'}}{dk'}$, \textit{i.e} $\frac{d^{2}\mathrm{D}_{k'}}{dk'^{2}}=0$ to get the value of $k_{0}$ which gives :
\begin{equation}
k_{0}(E) = 0.39 k_{E} = 0.39 \sqrt{\frac{2 m^{*}}{\hbar^{2}}} E^{1/2}\frac{\Gamma_{k'}(E)}{(|E-E_{k'}|)^{2}}
\end{equation}
We can now write Equ.S5 for $k=0$ where $F(k,k_{0})=2k_{0}$ as :
\begin{equation}
\Gamma_{0} = 0.02 M^{2} \sqrt{\frac{2m^{*}}{\hbar^{2}}} E^{1/2} \left\{\mathrm{N}_{pop} \mathrm{D}_{0}(E+E_{pop})+(\mathrm{N}+1) \mathrm{D}_{0}(E-E_{pop}) \right\}
\end{equation}

%\begin{tocentry}
%\centering
%\includegraphics[width=0.85 \textwidth]{fig_toc.jpg}

%\end{tocentry}

%\section*{Acknowledgement}

\eject

\bibliographystyle{achemso}
 
\bibliography{Ref1}

\end{document}